\documentstyle[12pt]{article}
\newcommand{\be}{\begin{equation}}
\newcommand{\ee}{\end{equation}}
\newcommand{\ba}{\begin{eqnarray}}
\newcommand{\ea}{\end{eqnarray}}

\begin{document}
\begin{center}
{\bf $SUSY-$APPROACH FOR INVESTIGATION OF TWO-DIMENSIONAL QUANTUM
MECHANICAL SYSTEMS}\footnote{Talk given at the International
Conference "Progress in Supersymmetric Quantum Mechanics",
Valladolid (Spain), July 15-20, 2003.} \vspace{0.2cm} \\{\bf
Mikhail V.Ioffe \footnote{E-mail: m.ioffe@pobox.spbu.ru}\\
\vspace{0.2cm} Department of Theoretical
Physics, St.-Petersburg State University,\\ 198504
Sankt-Petersburg, Russia }
\end{center}
\vspace{0.1cm}
\hspace*{0.5in}
\begin{minipage}{5.0in}
{\small Different ways to incorporate two-dimensional systems,
which are not amenable to separation of variables, into the
framework of Supersymmetrical Quantum Mechanics (SUSY QM) are
analyzed. In particular, the direct generalization of
one-dimensional Witten's SUSY QM is based on the supercharges of
first order in momenta and allows to connect the eigenvalues and
eigenfunctions of two scalar and one matrix Schr\"odinger
operators. The use of second order supercharges leads to
polynomial supersymmetry and relates a pair of scalar
Hamiltonians, giving a set of such partner systems with almost
coinciding spectra. This class of systems can be studied by means
of new method of $SUSY-$separation of variables, where
supercharges {\bf allow} separation of variables, but Hamiltonians
{\bf do not}. The method of shape invariance is generalized to
two-dimensional models to construct purely algebraically a chain
of eigenstates and eigenvalues for generalized Morse
potential models in two dimensions.}
\end{minipage}
\vspace*{0.1cm}

\section*{\bf 1. Introduction}
\vspace*{0.1cm}
\hspace*{3ex} Supersymmetric Quantum Mechanics \cite{witten}, \cite{review}
is an interesting framework
to analyze nonrelativistic quantal problems. In particular, it
allows to investigate the spectral properties of a wide class of
quantum models as well as to generate new systems with given
spectra. SUSY QM gives new insight into the problem of
spectral equivalence of Hamiltonians, which historically was
constructed as Factorization Method in Quantum Mechanics
\cite{infeld} and as Darboux-Crum transformations in Mathematical
Physics \cite{darboux}.

During last two decades SUSY QM became an important and popular
tool to study a wide variety of quantum systems (see the
list of reports presented to this Conference). It is easy to note
that the main stream of the development in SUSY QM concerned
one-dimensional models. Though the variety of
multi-dimensional (especially two- and three-dimensional) problems
is much wider and practically important, much less attention has
been given in the literature to the study of these models in
SUSY QM. Thus the future progress seems to be mostly connected with
investigation of its multi-dimensional generalizations. The main aim
of this paper (based on the talk at the Conference) is to summarize
different results of previous investigations of two-dimensional SUSY QM.

The paper is organized as follows. Two-dimensional generalization
of the conventional Witten's formulation of SUSY QM is formulated
in Section 2. In Section 3 the two-dimensional SUSY QM with
the supercharges of second order in derivatives is presented.
Section 4 contains a new, supersymmetric, approach for
investigation of two-dimensional models, which {\bf are not amenable} to
separation of variables. This method is based on the second order
supercharges introduced in the previous Section, and it gives a
new opportunity to reduce the problem to the one-dimensional ones.
Thus we obtain the specific method of $SUSY-${\bf separation of variables.}
Section 5 is devoted to the generalization of the well known
notion of shape invariance onto the two-dimensional models. Some new aspects
typical for two-dimensional shape invariance are investigated.
In Section 6
the deformation of SUSY QM algebra for models of Sections 3-5 is
described.

\section*{\bf 2. Direct two-dimensional generalization of
the conventional Witten's SUSY QM}
\hspace*{3ex} The conventional one-dimensional SUSY QM was
proposed by E.Witten \cite{witten}. It is characterized by the
simplest realization of SUSY algebra:
\ba
 \{\hat Q^+, \hat Q^-\} &=& \hat H;\quad
 (\hat Q^+)^2 =(\hat Q^-)^2 = 0;\quad
 [\hat H ,\hat Q^{\pm}] = 0, \label{hq}
\ea
where the superHamiltonian $\hat H$ is a diagonal matrix $\hat
H=diag(h^{(0)}, h^{(1)}) $, with $h^{(i)}=-\partial^2 +
V^{(i)}(x),\quad \partial\equiv d/dx,$ and supercharges
$\hat Q^{\pm}$ are off diagonal with elements $q^{\mp}\equiv
\pm\partial + \partial W(x)$ of first order in derivatives with
superpotential $W(x).$ In terms of components the
(anti)commutation relations of SUSY algebra (\ref{hq}) mean,
respectively, the
factorization of Hamiltonians, nilpotent structure of supercharges
and intertwining of $h^{(i)}$ by $q^{\pm}.$ The
superpotential $W(x)$ is defined by an arbitrary (possibly
non-normalizable) solution $\Psi (x)\equiv\exp{(-W(x))}$ of the
Schr\"odinger equation $h^{(0)}\Psi (x) = \epsilon \Psi (x) $ with
$\epsilon \leq E_0^{(0)}.$ If this solution is nodeless one has almost
coinciding spectra of $h^{(0)}$ and $h^{(1)}$ or, equivalently,
double degeneracy of the energy spectrum of $H$.

The direct multi-dimensional generalization of the construction
above was built in \cite{jetp} both by
direct extension \cite{abe} of one-dimensional formulas and
using \cite{abei} the
superfield approach of Quantum Field Theory. We call it as
"direct" since it retains both all relations of superalgebra
(\ref{hq}) and the first order form of the
components of supercharges.

Here we restrict ourselves to the particular case of {\bf two
space dimensions} $\vec x = (x_1, x_2).$ The precise formulas for
the $4\times 4$ superHamiltonian and supercharges are then the
following:
\ba
\hat H = \left(
\begin{array}{ccc}h^{(0)}(\vec x)&0&0\\ 0& h^{(1)}_{ik}(\vec
x)&0\\0&0&h^{(2)}(\vec x)
\end{array} \right);\quad i,k=1,2;\quad \hat Q^+ = (\hat Q^-)^{\dagger} =
\left(
\begin{array}{cccc}0&0&0&0\\
q_1^-&0&0&0\\q_2^-&0&0&0\\0&p_1^+&p_2^+&0
\end{array} \right);
 \label{2ham}
\ea
where two scalar Schr\"odinger operators $h^{(0)}, h^{(2)}$
and $2\times 2$ matrix Schr\"odinger operator $h^{(1)}_{ik}$ are
expressed in {\it quasifactorized} form in terms of the components
of supercharges:
\ba
h^{(0)}&=&q_l^+q_l^- = -\partial_l^2 +
V^{(0)}({\vec x}) = -\partial_l^2 + (\partial_lW(\vec x))^2 -
\partial_l^2W(\vec x);\,\, \partial_l^2 \equiv
\partial_1^2 + \partial_2^2; \nonumber\\
h^{(2)}&=&p_l^+p_l^-=-\partial_l^2 +
V^{(2)}({\vec x}) = -\partial_l^2 + (\partial_lW(\vec x))^2 +
\partial_l^2W(\vec x);\nonumber\\
h^{(1)}_{ik}&=&q_i^-q_k^+ + p_i^-p_k^+ = -\delta_{ik}
\partial_l^2 + \delta_{ik}((\partial_lW(\vec x))^2 - \partial_l^2W(\vec x))
+ 2\partial_i\partial_k W(\vec x).\nonumber
\ea

The components of supercharges $q_l^{\pm}, p_l^{\pm}$ are
again of first
order and depend on the two-dimensional superpotential $W(\vec x):$
\be
q_l^{\pm}\equiv \mp\partial_l + \partial_l W(\vec x);\quad
p_l^{\pm}\equiv \epsilon_{lk}q_k^{\mp}.\label{2p}
\ee

Two-dimensional $4\times 4$ superHamiltonian and supercharges
(\ref{2ham}) realize the same conventional SUSY QM algebra
(\ref{hq}). In components, the commutation relations in (\ref{hq})
are expressed as the intertwining relations between matrix
$h_{ik}^{(1)}$ and $h^{(0)}, h^{(2)}$ by operators $q_l^{\pm},
p_l^{\pm}$:
\be
h^{(0)}q_i^+=q_k^+h_{ki}^{(1)};\quad h_{ik}^{(1)}q_k^- =
q_i^-h^{(0)}; \quad h^{(1)}_{ik}p_k^-=p_i^-h^{(2)} ;
\quad p_k^+h^{(1)}_{ki} = h^{(2)}p_i^+ .\label{h0h1}
\ee

The energy spectra of $h^{(0)}$ and $ h^{(2)}$ are in general
different, but the intertwining relations (\ref{h0h1}) provide the
equivalence of energy spectra between a pair of two scalar
Hamiltonians $h^{(0)}$, $ h^{(2)}$ and $2 \times2$ matrix
Hamiltonian $h^{(1)}_{ik}.$ The "equivalence" means coincidence of
spectra up to zero modes of operators $q_l^{\pm}, p_l^{\pm}.$ Thus
the supersymmetry (supersymmetric transformation) allows to reduce
the solution of matrix Schr\"odinger problem with the Hamiltonian
$h^{(1)}_{ik}$ to solution of a couple of scalar Schr\"odinger
problems $h^{(0)}$, $ h^{(2)}.$ Due to the same intertwining
relations the vector wave functions of matrix Hamiltonian
$h_{ik}^{(1)}$ are also connected (up to a normalization factor)
with the scalar wave functions of scalar Hamiltonians $h^{(0)},
h^{(2)}:$
\ba
\Psi^{(1)}_i(\vec x; E) &=& q^-_i\Psi^{(0)}(\vec x;
E);\quad i=1,2\quad \Psi^{(0)}(\vec x; E) = q^+_i\Psi^{(1)}_i(\vec
x; E)\nonumber\\
\Psi^{(1)}_i(\vec x; E) &=&
p^-_i\Psi^{(2)}(\vec x; E)\quad\Psi^{(2)}(\vec x; E) =
p^+_i\Psi^{(1)}_i(\vec x; E).\nonumber
\ea

The Schr\"odinger operators with matrix
potential are not something very exotic in Quantum Mechanics. In
particular, the described two-dimensional generalization of SUSY QM
was successfully used \cite{pauli} to investigate the spectra of
Pauli operator for fermion in external electromagnetic fields.
Nevertheless, considering this rather good-looking construction, one question
seems to be natural: is it
possible to perform supersymmetric transformations in
two-dimensional case avoiding any matrix Hamiltonians?

\section*{\bf 3. Second order supercharges in two-dimensional SUSY QM}
\hspace*{3ex}The main idea, which could allow us
to get rid of matrix components of superHamiltonian, is to explore
the supercharges of second order in derivatives. For the first
time such supercharges of higher orders in momenta were proposed
for the one-dimensional situation in \cite{ais} (see also
\cite{acdi}, \cite{hsusy}) leading to the
polynomial deformation of SUSY algebra (see Section 6). In general, this approach
implies a deformation of one relation of SUSY algebra (\ref{hq})
only, namely of the (quasi)factorization, but keeping unchanged
the nilpotency of $\hat Q^{\pm}$ and the intertwining relations.
The last seem to be the most important ingredient of SUSY methods
in QM.

The simplest variant of second order supercharges $q^{\pm}$ - of
the so called reducible \cite{acdi} form - gives uninteresting
result in two-dimensional case: the intertwined
partner Hamiltonians differ by a trivial constant only, and both
of them admit the separation of variables (see details in
\cite{pl}, \cite{tmf}). By this reason here we will be interested in general
irreducible second order components of supercharges:
\ba
q^+ = g_{ik}{(\vec x)}\partial_i \partial_k + C_i(\vec x )\partial_i +
B(\vec x);\quad q^-\equiv (q^+)^{\dagger}.
\label{gench}
\ea

The important question we have to investigate now concerns the
existence of Hamiltonians
\be
h^{(i)}=-\Delta^{(2)}+V^{(i)}(\vec x );\quad i=1,2;\quad
\Delta^{(2)}\equiv \partial_l\partial_l , \label{hi}
\ee
which satisfy the intertwining relations with $q^{\pm}$ of the form
(\ref{gench}):
\be
h^{(1)}q^+=q^+h^{(2)};\quad q^-h^{(1)}=h^{(2)}q^-. \label{intertw}
\ee

The first consequence of (\ref{intertw}) restricts essentially the
possible "metrics" $g_{ik}(\vec x) $ by $\quad
\partial_lg_{ik} + \partial_ig_{lk}+\partial_kg_{il}=0 $
with solutions:
\be
g_{11} =\alpha x_2^2 + a_1 x_2 +  b_1;
\,\,g_{22} = \alpha x_1^2 + a_2 x_1 + b_2; \,\,g_{12}
=-\frac{1}{2}(2\alpha x_1 x_2 + a_1 x_1 + a_2 x_2) +
b_3. \nonumber
\ee
This has to be taken into account in rewriting \cite{pl}
the intertwining relations (\ref{intertw}) in components:
\ba
&&\partial_iC_k(\vec x ) +
\partial_kC_i(\vec x) + \Delta^{(2)} g_{ik}(\vec x) - (V^{(1)}(\vec x) -
V^{(2)}(\vec x))g_{ik}(\vec x) = 0; \label{sys1}\\
&&\Delta^{(2)}C_i(\vec x) + 2\partial_iB(\vec x) + 2 g_{ik}(\vec x
)\partial_k V^{(2)}(\vec x) - (V^{(1)}(\vec x) - V^{(2)}(\vec x
))C_i(\vec x)=0;\label{sys2}\\ &&\Delta^{(2)} B(\vec x) +
g_{ik}(\vec x)\partial_k\partial_i V^{(2)}(\vec x) + C_i(\vec x
)\partial_i V^{(2)}(\vec x) - (V^{(1)}(\vec x) - V^{(2)}(\vec x
))B(\vec x) = 0.\label{sys3}
\ea

The nonlinear system of second order differential equations
(\ref{sys1}) - (\ref{sys3}) for unknown functions $C_i(\vec x),\,
B(\vec x),\, V^{(1)}(\vec x),\, V^{(2)}(\vec x) $ with constant
parameters $\alpha, a_i, b_i$ in $g_{ik}(\vec x)$ does not admit
the general solution, but one can look for its particular
solutions with concrete metrics $g_{ik}$ and some ansatzes for
unknown functions $C_i .$

In particular, the system (\ref{sys1}) - (\ref{sys3}) is
essentially simplified for the metrics of elliptic form
$g_{ik}(\vec x)\equiv \delta_{ik} .$ In this case all unknown
functions in (\ref{sys1}) - (\ref{sys3}) can be found analytically
in the general form \cite{pl}, but for all such solutions both Hamiltonians
$h^{(1)},\,h^{(2)}$ turn out to admit the $R-$separation
\cite{miller} of variables in parabolic, elliptic or
polar\footnote{The reducible second order supercharges just
correspond to separation of variables in polar
coordinates.} coordinates, i.e. this class of two-dimensional
problems can be reduced to two one-dimensional models.

Much more interesting situation appears for more complicated forms
of metrics. Thus the list of particular solutions of the system
(\ref{sys1}) - (\ref{sys3}) can be constructed analytically
\cite{pl}, \cite{tmf}, \cite{classic} for
the hyperbolic metrics $g_{ik}=diag(+1,-1).$ Indeed, for this metrics
a part of task of solution of (\ref{sys1}) - (\ref{sys3}) can be
made in general form since it is reduced to a simpler system,
which will be written in terms of coordinates $x_{\pm}
\equiv x_1\pm x_2\quad \partial_{\pm}\equiv\partial / \partial
x_{\pm}$ and $C_+ \equiv C_1 - C_2;\quad C_- \equiv C_1 + C_2.$
The general solution can be provided by solving the system:
\ba
&&\partial_-(C_- F) = -\partial_+(C_+ F);\label{first}\\
&&\partial_+^2 F =
\partial_-^2 F,\label{second}
\ea
where $C_{\pm}$ depend only on $ x_{\pm},$ respectively:
$C_{\pm} \equiv C_{\pm}(x_{\pm}).$ The function $ F, $ solution of
(\ref{second}), is represented as $ F=F_{1}(x_{+}+x_{-}) +
F_{2}(x_{+}-x_{-}).$ The potentials $ V^{(1),(2)}(\vec x)$ and the
function $ B(\vec x) $ are expressed in terms of
$F_1(2x_1),\,F_2(2x_2)$ and $C_{\pm}(x_{\pm}),$ solutions of
(\ref{first}):
\ba
V^{(1),(2)}&=&\pm\frac{1}{2}(C_+' +
C_-') + \frac{1}{8}(C_+^2 + C_-^2) + \frac{1}{4}\biggl( F_2(x_+
-x_-) - F_1(x_+ + x_-)\biggr);\label{potential}\\
B&=&\frac{1}{4}\biggl( C_+ C_- + F_1(x_+ +
x_-) + F_2(x_+ - x_-)\biggr), \nonumber
\ea
where $C'$ means derivative in its argument.

For lack of a regular procedure for solution of both
equations of the system
(\ref{first}), (\ref{second}), its particular solutions can
be found starting from
certain anzatses for functions $C_{\pm}(x_{\pm}), F(\vec x)$.

1) Let $C_- = 0,$ then from (\ref{first}) one obtains
$F=\phi (x_-) / C_+(x_+).$ After inserting into Eq.(\ref{second})
the separation of variables is possible, and particular solution
reads\footnote{Here and below $F_{1,2}$ are defined only
up to an arbitrary real constant:
$F_1\to F_1 + \epsilon ,\,\, F_2\to F_2 - \epsilon.$}:
\ba
C_+(x_+)& =& \frac{1}{\delta_1 \exp(\sqrt{\lambda}
\cdot x_+) + \delta_2 \exp(-\sqrt{\lambda}\cdot
x_+)};\nonumber\\
F_{1,2}(2x)& =& \delta_1\sigma_{1,2}
\exp(2\sqrt{\lambda} x) + \delta_2\sigma_{2,1}
\exp(-2\sqrt{\lambda} x),\nonumber
\ea
the Greek letters -- arbitrary
constants, depending on sign of $\lambda$ they may be real/complex.

2) Let $F(\vec x)$ allows also the factorization:
$F = F_+(x_+)\cdot F_-(x_-)$. Then from Eq.(\ref{first}):
\be
C_{\pm} = \frac{\nu_{\pm}}{F_{\pm}} \pm \frac{\gamma}{F_{\pm}}\int
\limits^{x_{\pm}}F_{\pm}dx'_{\pm},\nonumber
\ee
and there are two options to fulfill the
condition (\ref{second}), i.e. $F(\vec x) = F_1(2x_1)+F_2(2x_2)$:
\be
a)\quad F_{\pm}(x_{\pm}) =
\epsilon_{\pm}x_{\pm},\qquad b)\quad F_{\pm} = \sigma_{\pm}
\exp(\sqrt{\lambda}\cdot x_{\pm}) + \delta_{\pm}
\exp(-\sqrt{\lambda}\cdot x_{\pm}).\nonumber
\ee

Corresponding potentials can be found according to
Eq.(\ref{potential}), being similar to ones obtained in
\cite{hietarinta} in quite different approach.
Below some other solutions of (\ref{first}), (\ref{second})
will be built \cite{classic}.

3) Let us start now from the general solution of (\ref{first}):
\be
F = L\biggl(\int\frac{dx_+}{C_+} - \int\frac{dx_-}{C_-}\biggr)/
(C_+C_-).\label{10}
\ee
Then Eq.(\ref{second}) gives the
functional-differential equation for the functional $ L(A_+-A_-)$
with $A^{\prime}_{\pm}\equiv 1/C_{\pm}(x_{\pm}):$
\be
\biggl(\frac{A_+'''}{A_+'} - \frac{A_-'''}{A_-'}\biggr)L(A_+ -
A_-) + 3 (A_+'' + A_-'')L'(A_+ - A_-) + (A_+'^2 - A_-'^2)L''(A_+ -
A_-) = 0, \label{11}
\ee
where $ L^{\prime} $ denotes the
derivative of $ L $ with respect to its argument.
If we take functions $ A_{\pm} $ such that
$A_{{\pm}}'' = \lambda^2 A_{\pm},\,\lambda = \mbox{const},$
Eq.(\ref{11}) will become ordinary differential
equation for $L$ with independent variable $(A_+-A_-).$
It can be easily solved:
\ba
L(A_+ - A_-) = \alpha
(A_+ - A_-)^{-2} + \beta,\nonumber
\ea
where $A_{\pm}=\sigma_{\pm}exp(\lambda x_{\pm}) +
\delta_{\pm}exp(-\lambda x_{\pm})$ with $
\sigma_{+}\cdot\delta_{+} = \sigma_{-}\cdot\delta_{-}$ and $
\alpha , \beta $ - real constants. For $ \lambda^{2}>0, $
choosing $\sigma_{\pm}=-\delta_{\pm}=k/2$ or
$\sigma_{\pm}=+\delta_{\pm}=k/2,$
we obtain (up to an arbitrary shift in $ x_{\pm} $) two
particular solutions:
\ba
3a)\quad A_{\pm} = k \sinh (\lambda x_{\pm}),\quad 3b)\quad
A_{\pm} = k \cosh (\lambda x_{\pm}).\nonumber
\ea
Then (\ref{10}) leads to:
\ba
3a)\quad &&F_1(2x) = \frac{k_1}{\cosh^2(\lambda x)} + k_2\cosh(2\lambda x);\label{12}\\
&&F_2(2x) = \frac{k_1}{\sinh^2(\lambda x)} + k_2\cosh(2\lambda x);\quad
C_{\pm} = \frac{k}{\cosh(\lambda x_{\pm})},\quad
k\not=0,\nonumber\\
3b)\quad &&F_1(2x) = - F_2(2x) =
\frac{k_1}{\sinh^2(\lambda x)} + k_2\sinh^2(\lambda x),\quad
C_{\pm} = \frac{k}{\sinh(\lambda x_{\pm})},\quad
k\not=0.\label{13}
\ea
For $ \lambda^{2}<0 $ hyperbolic functions
must be substituted by trigonometric ones.

We have to remark that the case $\lambda^2=0,$ i.e. $A_{\pm}^{\prime\prime}=0,$
is not of interest, leading to trivial superpartners. However, choosing
in (\ref{13}) $\lambda\to 0, k,\, k_1,\, k_2^{-1}\,\to 0 $ simultaneously,
so that $\lambda^2\sim k_1\sim k_2^{-1}\sim k^2,$ we obtain the solution:
\be
F_1(2x) = - F_2(2x) = \tilde k_1 x^{-2} + \tilde k_2 x^2, \quad
C_{\pm} = \frac{\tilde k}{x_{\pm}}. \label{14}
\ee
One can check that (\ref{first}) is also satisfied by
\be
F_1(2x) = - F_2(2x) = k_1 x^2 + k_2 x^4, \quad C_{\pm} =
\pm\frac{k}{x_{\pm}}. \label{15}
\ee

4) Starting again from (\ref{10}), it is convenient to pass on to
new variable functions $ C_{\pm}\equiv\pm f_{\pm}/f_{\pm}^{\prime}.$
Then $ F $ in (\ref{10}) is represented
in the form $ F=U(f_{+}f_{-})f_{+}^{\prime}f_{-}^{\prime}$ with an
arbitrary\footnote{Due to Eq.(\ref{second}), the function $F$ should
be additionally representable in the form $F = F_1(2x_1)+F_2(2x_2)$.}
function $ U.$ After substitution in (\ref{second}) one obtains
the functional-differential equation:
\ba
(f_+'^2 f_-^2 - f_+^2 f_-'^2)U''(f) + 3 f
\biggl(\frac{f_+''}{f_+} - \frac{f_-''}{f_-}\biggr)U'(f) +
\biggl(\frac{f_+'''}{f_+'} - \frac{f_-'''}{f_-'}\biggr)U(f) = 0,
\quad f\equiv f_+f_-.\nonumber
\ea
For particular form of functions $f_{\pm}=\alpha_{\pm}exp(\lambda x_{\pm}) +
\beta_{\pm}exp(-\lambda x_{\pm}),$ this equation becomes
an ordinary differential equation for $U$ with independent
variable $f.$ Its solution is $ U=a+4bf_{+}f_{-} $
$( a,b -$real constants). Then functions
\ba
F_1(x) &=&
k_1(\alpha_+\alpha_-\exp(\lambda x) + \beta_+\beta_-\exp(-\lambda
x)) + k_2(\alpha_+^2\alpha_-^2\exp(2\lambda x) +
\beta_+^2\beta_-^2\exp(-2\lambda x)),\nonumber\\
-F_2(x) &=&
k_1(\alpha_+\beta_-\exp(\lambda x) + \beta_+\alpha_-\exp(-\lambda
x)) +
 k_2(\alpha_+^2\beta_-^2\exp(2\lambda x) +
\beta_+^2\alpha_-^2\exp(-2\lambda x)),\nonumber\\
C_{\pm} &=& \pm
\frac{\alpha_{\pm}\exp(\lambda x_{\pm}) + \beta_{\pm}\exp(-\lambda
x_{\pm})} {\lambda(\alpha_{\pm}\exp(\lambda x_{\pm}) -
\beta_{\pm}\exp(-\lambda x_{\pm}))} \label{16}
\ea
(with $k_1\equiv a\lambda^2, \, k_2\equiv 4b\lambda^2$)
are real solutions of (\ref{first}), (\ref{second}), if $
\alpha_{\pm}, \beta_{\pm} $ are real for $ \lambda^{2}>0,$ and $
\alpha_{\pm}=\beta_{\pm}^{*} $ for $ \lambda^{2}<0.$

5) To find
a next class of solutions it is useful to rewrite (\ref{first}) in
terms of $ x_{1,2}:$
\ba
(F_1(2x_1) + F_2(2x_2))\partial_1(C_+ +
C_-) + F_1'(2x_1)(C_+ + C_-) + F_2'(2x_2)(C_+ - C_-) = 0.\nonumber
\ea
Among known particular solutions the most compact one is:
\be
C_+(x) = C_-(x) = a x^2 + c,\quad F_1(2x_1) = 0,\quad F_2(2x_2) =
\frac{b^2}{x_2^2}.\label{18}
\ee

After inserting these solutions (\ref{12}) - (\ref{18}) into the
general formulas (\ref{potential}), one obtains
the analytical expressions for potentials. Their explicit form
can be found in \cite{classic}.

The additional class of particular solutions of the system
(\ref{sys1}) - (\ref{sys3}) obtained for the case of degenerate
metrics $g_{ik}=diag(1,0)$ can be found also in \cite{classic}.

\section*{\bf 4. $SUSY-$ separation of variables}
\hspace*{3ex}From the very beginning in this paper
we are interested in two-dimensional quantum systems, which {\it
are not amenable} to separation of variables. The supersymmetric
approach, namely the intertwining relations (\ref{intertw}),
allows to formulate some specific supersymmetric alternative to
the conventional notion of separation (including the so-called
$R-$separation \cite{miller}) of variables. The main idea \cite{two}
is to consider such particular class of solutions of intertwining
relations (\ref{intertw}), when the components of supercharge
$q^{\pm}$ {\it are amenable} to separation of variables but
Hamiltonians $h^{(i)}$ {\it are not}. In this case the
Hamiltonians $h^{(i)}$ turn out to be partially solvable, or in
another terminology, quasi-exactly-solvable \cite{turbiner}. Both
terms above mean that a part of spectrum (and possibly of
corresponding eigenfunctions) of Hamiltonian is known. The crucial
ingredient of the approach \cite{two} is in the investigation of zero modes
of intertwining operators $q^{\pm}.$

The general scheme of the method is the following. Let us
suppose that $N+1$ normalizable zero modes of $q^+$ are known (for
example, due to separation of variables in $q^+$):
\be
q^+\Omega_n (\vec x)=0;\quad n=0,1,...,N;\qquad
q^+ \vec\Omega (\vec x)=0, \label{zero}
\ee
where $\vec\Omega (\vec x)$ is a column vector with components
$\Omega_n(\vec x).$ From the intertwining relations
(\ref{intertw}) one can see that the space of zero modes is closed
under the action of $h^{(2)},$ and therefore:
\be
h^{(2)}\vec\Omega (\vec x) = \hat C \vec\Omega (\vec x),
\nonumber
\ee
where $\hat C \equiv ||c_{ik}||$ is a
$c-$number $\vec x-$independent real matrix. If the matrix $\hat
C$ can be diagonalized
 by a real similarity transformation:
\be
 \hat B \hat C (\hat B)^{-1} = \hat\Lambda =
diag(\lambda_0,\lambda_1,...,\lambda_N), \label{diag}
\ee
the problem is reduced to a standard algebraic task within the zero
modes space:
\be
h^{(2)} (\hat B\vec\Omega (\vec x)) = \hat\Lambda(\hat B\vec\Omega
(\vec x)). \label{diagonal} \ee

It is not clear in advance, whether this general scheme is
realized practically? To put it differently, are there any
solutions $C_i(\vec x), B(\vec x)$ of the intertwining relations
(\ref{sys1}) - (\ref{sys3}), which give $q^+$ with separation of
variables?

To investigate this problem, it is useful \cite{two} to transform the
supercharge $q^+$ by the special similarity transformation, which
removes the terms linear in derivatives:
\be
\tilde q^+ = e^{(-\chi (\vec x))} q^+ e^{(+\chi (\vec x))} =
 \partial_1^2 -
\partial_2^2 + \frac{1}{4}(F_1(2x_1) + F_2(2x_2));
\,\, \chi (\vec x) = -\frac{1}{4}\bigl( \int
C_+(x_+)dx_+ + \int C_-(x_-)dx_- \biggr). \label{chi}
\ee

These new operators $\tilde q^+$ obviously obey the condition of
separation of variables realizing the first step of our
scheme of $SUSY-$separation of variables. Zero modes of $\tilde
q^+ $ can be found as linear superpositions of products of one
dimensional wave functions $\eta_n (x_1)$ and $\rho_n(x_2),$
satisfying Schr\"odinger equations (with $\epsilon_n$ - the
separation constants.):
\be
(-\partial_1^2
-\frac{1}{4}F_1(2x_1))\eta_n(x_1)=\epsilon_n\eta_n(x_1); \qquad
(-\partial_2^2
+\frac{1}{4}F_2(2x_2))\rho_n(x_2)=\epsilon_n\rho_n(x_2).
\label{etarho}
\ee
In analogy to (\ref{chi}), one can define operators
\be
\tilde h \equiv \exp(-\chi (\vec x))\, h^{(2)}\,\exp(+\chi (\vec
x))= -\partial_l^2+C_1(\vec x)\partial_1 - C_2(\vec x)\partial_2
-\frac{1}{4} F_1(2x_1) +\frac{1}{4}F_2(2x_2), \nonumber
\ee
and eigenfunctions of $\tilde q^+$ as:
\be
\tilde\Omega_n(\vec x)= \exp(-\chi (\vec x))\cdot\Omega_n(\vec x),
\label{omegatilde}
\ee
keeping however in mind that the
normalizability and orthogonality are not preserved automatically
due to non-unitarity of the similarity transformation.

Then using (\ref{etarho}) one can write:
\be
\tilde h\tilde\Omega_n(\vec x)=[2\epsilon_n +C_1(\vec x)\partial_1
-C_2(\vec x)\partial_2]\tilde\Omega_n(\vec x). \label{eigen}
\ee
It is
not evident from (\ref{eigen}), but the space spanned by functions
$\tilde\Omega_n(\vec x)$ is closed under the action of $\tilde h.$
It will be demonstrated explicitly in the concrete model below.

In contrast to (\ref{chi}), where variables are separated, no
separation for $\tilde h,$ which would make the two-dimensional
dynamics not-trivially reducible to one-dimensional dynamics. In
this regard we refer to this method \cite{two} for partial solvability as to
{\bf $SUSY-$separation of variables}.

The scheme of $SUSY-$ separation of variables formulated above can
be used for arbitrary models satisfying the intertwining relations
(\ref{intertw}). The list of such models is already rather long,
and it may increase in future, but it is very important to check
the applicability of the scheme on the concrete model where the
explicit solutions can be constructed. Actually, it means that
solutions of two one-dimensional problems (\ref{etarho}) can be
found analytically. Below we briefly describe such a model -
generalized two-dimensional Morse potential.

Among the solutions \cite{classic}
of the system (\ref{first}), (\ref{second}) we focus attention on the
particular case with a specific
choice of parameters and $A>0, \alpha >0, a$--real constants
\footnote{For the complexification of the model see \cite{pseudo}}:
\ba
C_+&=&4a\alpha;\quad C_-=4a\alpha\cdot\coth \frac{\alpha x_-}{2};
\label{cpm}\\
f_i(x_i)&\equiv & \frac{1}{4}
F_i(2x_i)=-A\biggl(e^{-2\alpha x_i} - 2 e^{-\alpha
x_i}\biggr);\quad i=1,2;
\label{f2}\\
V^{(1),(2)}&=&
\alpha^2a(2a \mp 1)\sinh^{-2}\biggl(\frac{\alpha x_-}{2} \biggr) +
4a^2\alpha^2 + A \biggl[e^{-2\alpha
x_1}-2 e^{-\alpha x_1} + e^{-2\alpha x_2}-2 e^{-\alpha
x_2}\biggr].\label{morse}
\ea

One easily recognizes in (\ref{morse}) a sum of two Morse
potentials plus a hyperbolic singular term which prevents to apply
the {\it conventional} methods of separation of variables.
These singular terms can be both attractive, for $|a| > \frac{1}{2},$ or
one repulsive and one attractive, for $|a| <
\frac{1}{2}.$ The parameter $a$ will be further constrained by the
condition that the strength of the attractive singularity at
$x_-\to 0$ should not exceed the well known bound $-1/(4x_-^2).$

The normalizable functions $\tilde\Omega_n(\vec x)$ (and $\Omega_n(\vec
x )$) can be constructed from the well known \cite{landau}
normalizable solutions of (\ref{etarho}) with $\epsilon_n < 0 :$
\be
\tilde\Omega_n(\vec x) = \exp(-\frac{\xi_1+\xi_2}{2}) (\xi_1\xi_2)^{s_n}
F(-n, 2s_n +1; \xi_1) F(-n, 2s_n +1; \xi_2), \label{hyper}
\ee
where $F(-n, 2s_n +1; \xi) $ is the standard degenerate
(confluent) hypergeometric function, reducing to a polynomial for
integer $n,$ and
\be
\xi_i\equiv \frac{2\sqrt{A}}{\alpha}\exp(-\alpha x_i); \quad
s_n=\frac{\sqrt{A}}{\alpha}-n-\frac{1}{2}
> 0; \quad
\epsilon_n=-A\biggl[1-\frac{\alpha}{\sqrt{A}}(n+\frac{1}{2})\biggr]^2.
\label{epsilon}
\ee
The number $(N+1)$ of normalizable zero modes
(\ref{hyper}) is determined by the inequality $s_n>0.$

The condition of normalizability of zero modes $\tilde\Omega_n(\vec x
),$ together with the absence of the "fall to the centre", leads
\cite{two} to the following two ranges of parameters:
\ba
&&a \in
(-\infty,\, -\frac{1}{4}-\frac{1}{4\sqrt{2}}); \quad s_n =
\frac{\sqrt{A}}{\alpha}-n-\frac{1}{2} > -2a >0. \label{region1}\\
&&a \in ( -\frac{1}{4} ,\,\frac{1}{4} ); \quad s_0>2(|a|+1).
\label{region2}
\ea
In Sections 4 and 5 only the region
(\ref{region1}) will be considered. Inequalities (\ref{region1})
can be satisfied by the choice of $a$ and $A,$
and/or by suitable restriction on the number $N$ of
zero modes $\Omega_n(\vec x).$

Analysis of the action of $\tilde h$ in (\ref{eigen}) gives that
the matrix $\hat C$ is of triagonal form. It can be diagonalized
explicitly by a similarity transformation, and the eigenvalues
$E_k$ of $h^{(2)}$ coincide with its (all different and nonzero)
diagonal elements:
\be
E_k=c_{kk}=-2(2a\alpha^2s_k-\epsilon_k).
\label{energy}
\ee
The resulting eigenfunctions of $h^{(2)}$ are
obtained (see Eqs.(\ref{diag}), (\ref{diagonal})) from the
constructed zero modes $\tilde\Omega_n(\vec x)$ and the similarity
transformation $\hat B :$
\be
\Psi_{N-n}(\vec x) = \Sigma_{l=0}^N b_{nl}\Omega_l(\vec x).
\label{psiomega}
\ee
For the algorithm of iterative construction
of coefficients $b_{nl}$ see \cite{two}. Thus the construction of
the set of eigenfunctions, which lie in the space of zero modes,
is completed.

These eigenfunctions $\Psi_k(\vec x)$ may be also used for
constructing more general eigenfunctions of $h^{(2)}$ via a
product ansatz:
\be
\Phi (\vec x)\equiv \Psi_k(\vec x)\cdot\Theta (\vec x).
\label{product}
\ee
Three such eigenfunctions based on $\Psi_0$
were constructed in \cite{two}. Within the bounds imposed
(\ref{region1}) only one of them is normalizable, though for the
region (\ref{region2}) all three are normalizable.

\section*{\bf 5. Shape invariance in two dimensions}
\hspace*{3ex}  In the previous Section we developed
the method which led to construction of the partially solvable
(quasi-exactly-solvable) two-dimensional models. Let us remind now
the well known {\it in one dimension} and very elegant method of
{\bf shape invariance} \cite{shape} usually associated with the
exactly solvable one-dimensional systems. Our aim here is to
generalize the idea of shape invariance to the two-dimensional
case \cite{two}.

Refering readers to the original paper \cite{shape} and reviews
\cite{review} for the detail discussion of standard
one-dimensional shape invariance method, let us list its main
steps only:
\be
\tilde H(x;a)=H(x;\bar a) + {\cal R}(a);\quad \bar a=\bar a(a)
\label{shape1}
\ee
where $ {\cal R}(a)$ is a ($c$-number) function
of $a.$ The absence of spontaneous breaking of supersymmetry for
all values of $a$ implies that the lowest eigenvalue $E_0(a)$ of
$H(a)$ vanishes and the corresponding eigenfunctions $\Psi_0(a)$
are normalizable zero modes of $Q^+(a).$

The intertwining relations
$Q^-(x;a)\tilde H(x;a)=H(x;a)Q^-(x;a)$ with the
standard first order supercharge allow in this case to solve the
entire spectral problem for $H(x;a).$ The crucial steps are as
follows.
\be
H(x;\bar a)\Psi_0(x;\bar a)= E_0(\bar a)\Psi_0(x;\bar
a)=0; \quad \tilde H(x;a)\Psi_0(x;\bar a)={\cal R}(a)\Psi_0(x;\bar
a). \label{sch}
\ee
It is important to remark that $\Psi_0(x;\bar
a)\equiv \tilde\Psi_0(x;a)$ has no nodes and therefore is the
ground state wave function of $\tilde H(x;a).$ Then
\be
H(x;a)\biggl[ Q^-(x;a)\Psi_0(x;\bar a)\biggr]= {\cal R}(a)\biggl[
Q^-(x;a)\Psi_0(x;\bar a)\biggr]. \label{R}
\ee
Provided $\biggl[
Q^-(xa)\Psi_0(x;\bar a)\biggr]$ is normalizable, we have generated
an excited state of $H(x;a),$ and thus ${\cal R}(a)$ is naturally
positive. It is clear that these steps can be repeated up to the
last one, where the resulting wave function $\Psi$ will no more be
normalizable.

It is also clear that the isospectrality of $H(x;a)$ and $\tilde
H(x;a)$ (up to the only zero mode $\Psi_0(x;a) )$ implies that
there is no eigenvalue of $H(x;a)$ between zero and the ground
state energy $\tilde E_0(a)$ of $\tilde H.$ This observation leads
to a proof that after suitable iterations one gets the entire
spectrum of $H(x;a).$ This method is referred as algebraic
solvability (or complete solvability) by shape invariance in
one-dimensional SUSY QM.

To proceed to the formulation of two-dimensional shape invariance,
we start from the relatively simple two-dimensional case of
systems with conventional separation of variables: $$ H(\vec x
)=H_1(x_1)+H_2(x_2);\quad \vec x=(x_1,x_2). $$ Now suppose that
$H_1$ and $H_2$ both are shape invariant: \be \tilde H_i(x_i;
a_i)=H_i(x_i;\bar a_i)+{\cal R}_i(a_i)\quad \leftrightarrow \quad
\tilde H(\vec x;{\bf a})=H(\vec x;{\bf \bar a})+{\cal R}({\bf
a});\, {\bf a}\equiv (a_1,a_2).\ee In order to realize a
nontrivial intertwining relations for $H, \tilde H$ one can
consider factorized supercharges of second order written as
products of first order supercharges:
\be
Q^{\pm}=Q_1^{\pm}\cdot Q^{\pm}_2; \quad Q^{\pm}_i= \mp\partial_i +
W_i(x_i). \label{sepcharge}
\ee

There is the considerable difference with respect to the
one-dimensional case. The crucial reason is that the space of zero
modes of supercharges becomes now of higher dimensionality
including the products of one-dimensional zero modes of the first
Hamiltonian times all states of the second Hamiltonian and vice
versa. While iterations are again obviously possible, it is clear
that one can not argue about the entire solvability of the
spectral problem, because in general many zero modes of
(\ref{sepcharge}) exist. Their number depends on the confining
properties of $H_1$ and $H_2.$ For example, in a case of
oscillator-like potentials this number becomes infinite, and they
are distributed over the whole spectrum. In this case only partial
solvability of $H$ can be achieved by the choice of
(\ref{sepcharge}) and shape invariance. Of course, one can solve
such trivial models by separate use of $Q^{\pm}=Q^{\pm}_i,$ which
allows to solve the entire spectrum of two-dimensional model in
terms of the one-dimensional ones.

Let us suppose to have two-dimensional system (without separation
of variables) with a Hamiltonian $H,$ which is related to $\tilde
H$ by (\ref{shape1}). For simplicity (in general, there is no
connection between the dimensionality of the Schr\"odinger
equation and the dimensionality of the parameter manifold), we
assume that shape invariance is realized with one parameter $a.$
Two-dimensional SUSY QM does not identify here zero modes of
$Q^{\pm}$ with the ground state of the Hamiltonian. Thus one has
to repeat the steps (\ref{shape1}) - (\ref{R}) above by taking
into account $E_0(a)\neq 0.$ In order to make our discussion more
explicit we will from now on refer explicitly to the model
(\ref{cpm}) - (\ref{morse}) $(H\equiv h^{(2)}),\, \tilde H\equiv h^{(1)}$
with the parameter $a$ being bound to (\ref{region1}).

First of all we observe that this model is indeed shape invariant
(the infinite domain given by (\ref{region1}) allows iterations of
(\ref{shape1})):
\be
\bar a = a-\frac{1}{2};\quad {\cal
R}(a)=\alpha^2(4a-1) ) \label{bara}
\ee
The starting point is to write (\ref{R}):
\be
H(\vec x;a)\biggl[ Q^-(\vec x;a)\Psi_0(\vec x;a-\frac{1}{2})
\biggr] = \biggl(E_0(a-\frac{1}{2}) + {\cal
R}(a)\biggr)\cdot\biggl[ Q^-(\vec x;a)\Psi_0(\vec x;a-\frac{1}{2})
\biggr], \label{RR}
\ee
where $E_0(a)$ and $\Psi_0(\vec x;a)$ not
to be identified with ground state. Thus we have constructed the
new eigenstate and eigenvalue of $H(\vec x;a),$ provided $Q^-(\vec
x; a)\Psi_0(\vec x;a-\frac{1}{2})$ is normalizable. Note that the
eigenvalue $\biggl(E_0(a-\frac{1}{2}) + {\cal R}(a)\biggr)$ is
larger than $E_0(a)$ with the bounds of (\ref{region1}).

It is interesting that the energy of the first iteration of shape
invariance in (\ref{RR}) coincides precisely with the eigenvalue
of the additional solution $\Phi(\vec x)$ mentioned in the previous
Section: $$ E=E_{0}(a-\frac{1}{2})+ {\cal R}(a)=
\alpha^{2}[4a(1-s_{0})+(2s_{0}-1)]+2\epsilon_{0}.$$

The next iteration of shape invariance  will give:
\be
H(a)\biggl[ Q^-(a)Q^-(a-\frac{1}{2})\Psi_0(a-1) \biggr] =
\biggl(E_0(a-1) + {\cal R}(a-\frac{1}{2}) + {\cal
R}(a)\biggr)\cdot \biggl[ Q^-(a)Q^-(a-\frac{1}{2})\Psi_0(a-1)
\biggr], \label{RRR}
\ee
and the new eigenfunction $
Q^-(a)Q^-(a-\frac{1}{2})\Psi_0(a-1) $ can be written explicitly as
function of $ \vec x . $ Provided normalizability is ensured, one
can thereby construct a chain by successive iterations of
(\ref{RR}) and (\ref{RRR}), since $Q^-(a)$ has no normalizable
zero modes in (\ref{region1}). The end point of such a chain will
be given by non-normalizability of the relevant wave function.

Let us stress that though we illustrated both methods by the same
model, $SUSY-$separation of variables can be implemented
completely independently from shape invariance. For example, the
model considered in the range (\ref{region2}) admits the method of
$SUSY-$separation of variables, but shape invariance can not be
applied since the domain (\ref{region2}) is too small.

\section*{\bf 6. Polynomial algebra for two-dimensional SUSY QM}
\hspace*{3ex} In this last Section we will analyze
the deformation of conventional SUSY QM algebra (\ref{hq}) due to
introducing of second order components $q^{\pm}$ of supercharges
in Sections 3-5. It is obvious that keeping the intertwining
relations (\ref{intertw}) and the matrix structure of supercharges
$\hat Q^{\pm}$ we do not change two relations of SUSY algebra
(\ref{hq}), which express the supersymmetry of $\hat H$ and
nilpotency of $\hat Q^{\pm}$. But the third relation, which gives
the (quasi)factorization of the components of $\hat H,$ cannot be
fulfilled. In the one-dimensional case the anticommutator of $\hat
Q^{\pm}$ gives \cite{acdi} the operator of fourth order in
derivatives which can be represented as a second order polynomial
of the superHamiltonian $\hat H.$ The situation changes crucially
for two-dimensional systems of Sections 3-6, where in general a
new {\it diagonal} operator of fourth order appears:
\be
\hat R \equiv \{ \hat Q^+, \hat Q^- \} \label{RRRi}
\ee
This
operator obviously commutes with the superHamiltonian $\hat H$ due
to supersymmetry of $\hat H$ (intertwining relations
(\ref{intertw})). It is shown in \cite{tmf} that for Laplacian
metrics in (\ref{gench}) $g_{ik}=\delta_{ik}$ (where variables can
be separated) operator $\hat R$ can be reduced to the second order
symmetry operator $\tilde R$ up to a second order polynomial (with
constant coefficients) of $\hat H.$ But for all other metrics
$g_{ik},$ including hyperbolic and degenerate ones, the Theorem was
proved \cite{tmf}: {\it The symmetry operator} $\hat R$  {\it is
essentially of fourth order in derivatives, its order can not be
reduced}. The components $R^{(i)},\, i=1,2$ of $\hat R$ are the
symmetry operators of the components $h^{(i)}:\,\, [ h^{(i)},
R^{(i)}] = 0.$ The explicit expressions of $R^{(i)}$ for known
solutions of (\ref{sys1}) - (\ref{sys3}) can be written (see
\cite{lomi}). Thus all two-dimensional models of Section 3 are
{\bf integrable}\footnote{We note that integrability of the system does not
mean in general its (even partial) solvability.}.

As for generalized Morse model of Section 4, the quantum integral
of motion $R=Q^-Q^+$ gives zero on the eigenfunctions $\Psi_k(\vec
x)$ by construction, since they are zero modes of $Q^+.$ But by a
direct calculation one can check that all three additional
eigenfunctions $\Phi (\vec x)$ (see (\ref{product})) of $H$ are
simultaneously \cite{two} eigenfunctions (with nonzero
eigenvalues) of the symmetry operator $R.$ Thus they belong to a
system of common eigenfunctions of two Hermitian mutually
commuting operators $H$ and $R.$

\section*{\bf Acknowledgements}
The most part of results presented in this talk was obtained in
collaboration with my colleagues A.A.Andrianov, N.V.Borisov,
F.Cannata, M.I.Eides and D.N.Nishnianidze, to whom I am very
grateful for cooperation. I am indebted to Organizing Committee of
the Conference at Valladolid, especially to L.M.Nieto, J.Negro and
B.F.Samsonov, for invitation, support and their kind
hospitality at Valladolid . The participation at the Conference
was partially supported by RFFI (grants 02-01-00499 and
03-02-26770).

\end{document}